\documentclass{aa}
\usepackage{graphicx}

\begin{document}
%
\newcommand{\hi}{\ion{H}{i}~}
\newcommand{\hii}{\ion{H}{ii}~}
%


   \title{The \hi supershell GS061+00+51 and its neighbours} 

   \author{S. Ehlerov\' a \inst{1}
           \and
           J. Palou\v s \inst{1}
           \and
           W.K. Huchtmeier \inst{2}
           }

   \offprints{S. Ehlerov\'a }

   \institute{Astronomical Institute,
         Academy of Sciences of the Czech Republic,
         Bo\v cn\' \i \ II 1401, 141 31 Prague 4, Czech Republic \\
         \and 
         MPIfR, Auf dem H\"ugel 69, D-53121 Bonn, Germany \\
              }

   \date{Received 27 February 2001/ Accepted 17 May 2001}

   \titlerunning{The \hi supershell GS061+00+51 and its neighbours}
   \authorrunning{Ehlerov\'a et al.}

\abstract{
We describe \hi observations of a $4{\degr} \times 4{\degr}$ field
in the Milky Way centered on $l=61{\degr}$, $b=0{\degr}$ made by the
Effelsberg radiotelescope. The field contains one previously identified
\hi supershell, GS061+00+51 (Heiles, 1979); apart from it we find several
new structures. We also study the \hi distribution in the vicinity of
four \hii regions, S86, S87, S88 and S89. 
We confirm the existence of the shell GS061+00+51, and we find
that it has two smaller neighbours, spherical shells with 
$R_{\mathrm{sh}} \sim 30\ \mathrm{pc}$. We identify at least one more
regular shell at $v_{\mathrm{LSR}} = -18\ \mathrm{km s^{-1}}$; and 
one blown-out shell at  
$v_{\mathrm{LSR}} = -54\ \mathrm{km s^{-1}}$. In two cases we are
able to connect \hii regions with features in the \hi distribution
(S86 and S87), in two other cases no connection is found. Apart from
quite regular \hi shells we see a number of non-coherent objects,
which are probably a result of the turbulence in the interstellar
medium.
          }

   \maketitle

\section{Introduction} 

Turbulence creates in the interstellar medium (ISM) many structures, 
typically dense sheets, clumps and low-density holes. The majority of
these structures are transient. Many of them have an irregular, patchy 
appearance; however, some may look like ordinary, regular objects.

Another type of structures found in the ISM are \hi shells and holes. 
We agree with Walter \& Brinks 
(1999), that there is a difference between turbulent structures and \hi 
shells, at least in the sense that most turbulent 
structures show very little consistency if any in the position-velocity
(or position-position) space, while \hi shells do. This, of course, does not
mean, that the turbulent medium does not influence the shape and evolution
of \hi shells. It does, and as a first guess we may estimate that shapes
of shells in a turbulent medium will be more irregular than in a
smooth medium.

We observed a field in the galactic plane, which contains the supershell 
GS061+00+51 (Heiles, 1979). Our observations have four times 
higher resolution than the survey of Weaver \& Williams (1973) used
for the previous identification. Apart from the shell GS061+00+51 and its 
surroundings we study the rest of the datacube and try to identify new 
shells and shell-like structures.

In the observed field four optical \hii regions are known (S86, S87, S88 and
S89; Sharpless, 1959), at least one of them (S86) is connected to an OB
association (Vul OB1). The angular dimensions of the mentioned \hii regions
are greater than or comparable to the resolution of our observations, and
therefore we should be able to see their imprint in the \hi distribution.

\section{Observations, data reduction}

In 1997 (March--June) we observed a $4{\degr} \times 4{\degr}$
field in the Milky Way centered on $l=61{\degr}$, $b=0{\degr}$
with the 100~m radiotelescope in Effelsberg at the frequency
1.4~GHz (21~cm) of the neutral hydrogen line. The frequency 
switching mode was used. 
The bandwidth 1.56~MHz was split into 512 channels with the width 
of $\sim $3 kHz, or 0.64 $\mathrm{kms}^{-1}$.
The primary beamwidth of the Effelsberg radiotelescope at 21~cm is
9.4 arcmin, observations were made with a spacing of 4 arcmin
(the pixel size). Each spectrum was integrated for 15 s. The data
were calibrated using the standard S7 procedure (Kalberla et al.,
1982) and a linear baseline was subtracted. Observations were made in 
6 runs, each dataset was calibrated separately.
Data were not corrected for stray radiation, because we observed
a small field and were mostly interested in the differential effect 
of the observed emission against the background. Radio frequency 
interference may be present in the observational data. 

\begin{figure}
 \centering
 \includegraphics[angle=0,width=\hsize]{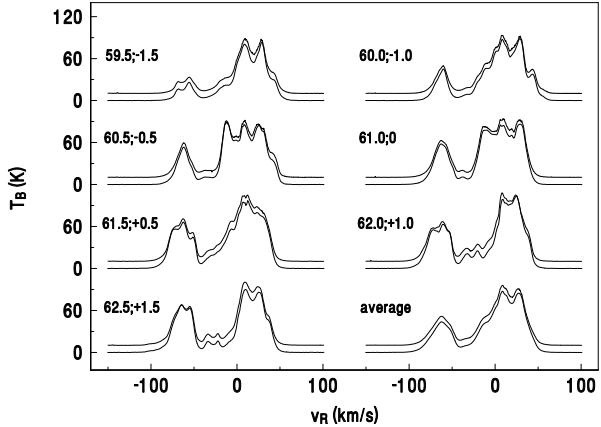}
 \caption{A comparison between the Leiden-Dwingeloo survey (thick
          line) and our observations (thin line).
          Our spectra are artificially offset by 10 K.}
 \label{fig1}
\end{figure}

To check our observations we compared them with data
from the Leiden-Dwingeloo \hi survey (Hartmann \& Burton, 1997),
which had a resolution of $0.5{\degr}$. To match the Dwingeloo beam 
we averaged spectra from 81 pixels, i.e. (36 arcmin)$^2$.
A comparison between our data and the Dwingeloo survey is shown in 
Fig. \ref{fig1}, where good agreement between the two
datasets is visible.

\section{Identification of \hi shells}

To be ranked among \hi shells, a structure must fulfill several
criteria:
\begin{itemize}
\item{it is either a hole -- i.e. a region of lower brightness
temperature, or a shell -- a sheet of higher temperature,
or both -- a hole encircled by a shell;} 
\item{it must be visible in several consecutive velocity
channels, gradually changing its size;}
\item{preferably, it should be expanding.}
\end{itemize}

The identification was made by eyes and this process is quite subjective. 
Both the eye and the criteria stated above have a bias towards regular 
(spherical) and compact structures, while less regular structures
are overlooked; the type of background (smooth vs. turbulent) also 
plays an important role. We suppose that the identified structures are real,
but we definitely missed some shells, especially in crowded or very 
turbulent regions.

Due to the size of the field, many structures are only partly visible,
and those we do not describe here (of the known \hi
shells GS064-0.1-97 (Heiles 1979) is seen in channel maps as a 
partial arc). Another previously known structure, the shell GS061+00+51, 
lies fully in the observed field.

\begin{figure}
 \includegraphics[angle=0,width=\hsize]{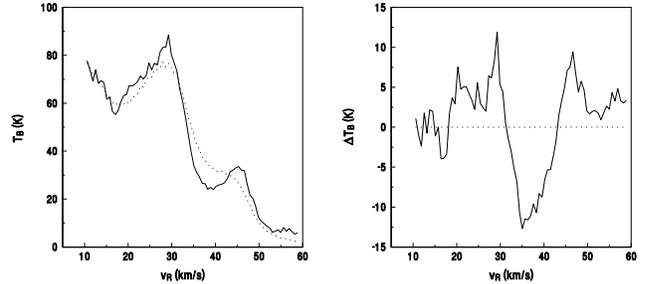}
 \caption{$\Delta T_{\mathrm{B}}$ graph of the shell GS59.9-1.0+38.
    The left panel shows the spectrum through the center of the
    shell (solid line) and the average spectrum of the surrounding
    area (dotted line). The right panel shows their difference
    $\Delta T_{\mathrm{B}}$, where both walls (receding and approaching)
    belonging to the structure and the hole (corresponding to the swept-up 
    region) are nicely recognizable.
         }
 \label{fig2}
\end{figure}

\subsection{$\Delta T_B$ graphs}

The spectrum of an \hi shell (in an ideal case) contains two peaks
corresponding to the intersection of the line of sight with dense walls and
a depression corresponding to the hole. These features are superimposed on the
spectrum emitted by the surrounding ISM. To disentangle the two contributions,
we subtract the background emission from the spectrum through the 
structure; the shell features should then appear. 
The column densities of the gas swept up into the wall 
can be --- under some assumptions on the shape and
dimensions of the shell --- transformed into the mass of the shell and the
initial volume density of the ISM. 

However, to determine the background spectrum is difficult,
first because of the 
unknowns in the velocity field and gas distribution which shape the spectrum,
secondly because of the turbulent character of the ISM. The simplest way 
to define the background is to take the average of the emission from a region
around the studied line of sight (in the case of studying the spectrum 
through the
\hi shell the region should contain the whole structure). 
This approach has its drawbacks, but at least it
smears out the small scale inhomogeneities.
When applied to artificial datacubes, we find, that often this 
method leads to a slight underestimation of the real values (lower
column densities of the swept-up gas and lower masses).
 
An important ``spoiler'' is the non-zero velocity dispersion of the gas, 
both in the shell and in the ISM. How important this effect is, depends
on the ratio between the velocity dispersion and the expansion velocity.
The line widths of the walls correspond to the real velocity dispersion
in the gas swept in the shell. The line width of the hole is not
so easy to classify and so we abstain from any deductions.

It is also possible to estimate masses purely from the dimensions of the shell
and an assumed (or estimated or fitted) density $n_0$ of the ISM. This 
approach, due to the variability of $n_0$ on many scales, does not lead 
to better or more reliable results. 

\subsection{Distances and energetics}

To calculate kinematic distances of shells we use the rotation curve 
of Wouterloot et al. (1990).

The total energy $E_{\mathrm{tot}}$ required to create the \hi shell 
is estimated using the Chevalier (1974) formula
\begin{equation}
    {E_{\mathrm{tot}} \over \mathrm{erg}} = 5.3 \times 10^{43}
    \left ({n_0 \over \mathrm{cm}^{-3}} \right )^{1.12}
    \left ({R_{\mathrm{sh}} \over \mathrm{pc}} \right )^{3.12}
    \left ({v_{\mathrm{exp}} \over \mathrm{kms}^{-1}} \right )^{1.4}
    \label{eq:cheval}
\end{equation}
where $n_0$ is the density of the ambient medium, $R_{\mathrm{sh}}$ the
radius of the shell and $v_{\mathrm{exp}}$ its expansion velocity.

\section{\hi shells: description}

\subsection{GS59.9-1.0+38}

\begin{table}[htb]
\begin{tabular}{lrrr}
\hline
~~ & FWHM & $N_{\mathrm{\hi}}$ & $m_{\mathrm{\hi}}$ \\
~~ & ($\mathrm{km s^{-1}}$) & ($\mathrm{cm^{-2}}$) & ($M_{\sun}$) \\
\hline
hole    & 7.1 & $1.5 \times 10^{20}$ & $2.9 \times 10^3$ \\
wall 1  & 3.2 & $4.6 \times 10^{19}$ & $5.6 \times 10^3$ \\
wall 2  & 2.6 & $3.7 \times 10^{19}$ & $4.5 \times 10^3$ \\
\hline
\end{tabular}
\caption{GS59.9-1.0+38:
FWHM is the width of the line, $\mathrm{N_{\hi}}$ is the column density 
of \hi swept up into the wall or missing in the hole derived from the 
spectrum through the centre of the shell, $m_{\mathrm{\hi}}$ 
is the mass of \hi swept up in walls or missing in the hole, 
assuming the radius of 35 pc.}
\label{table1}
\end{table}

\begin{figure*}
 \includegraphics[width=8cm]{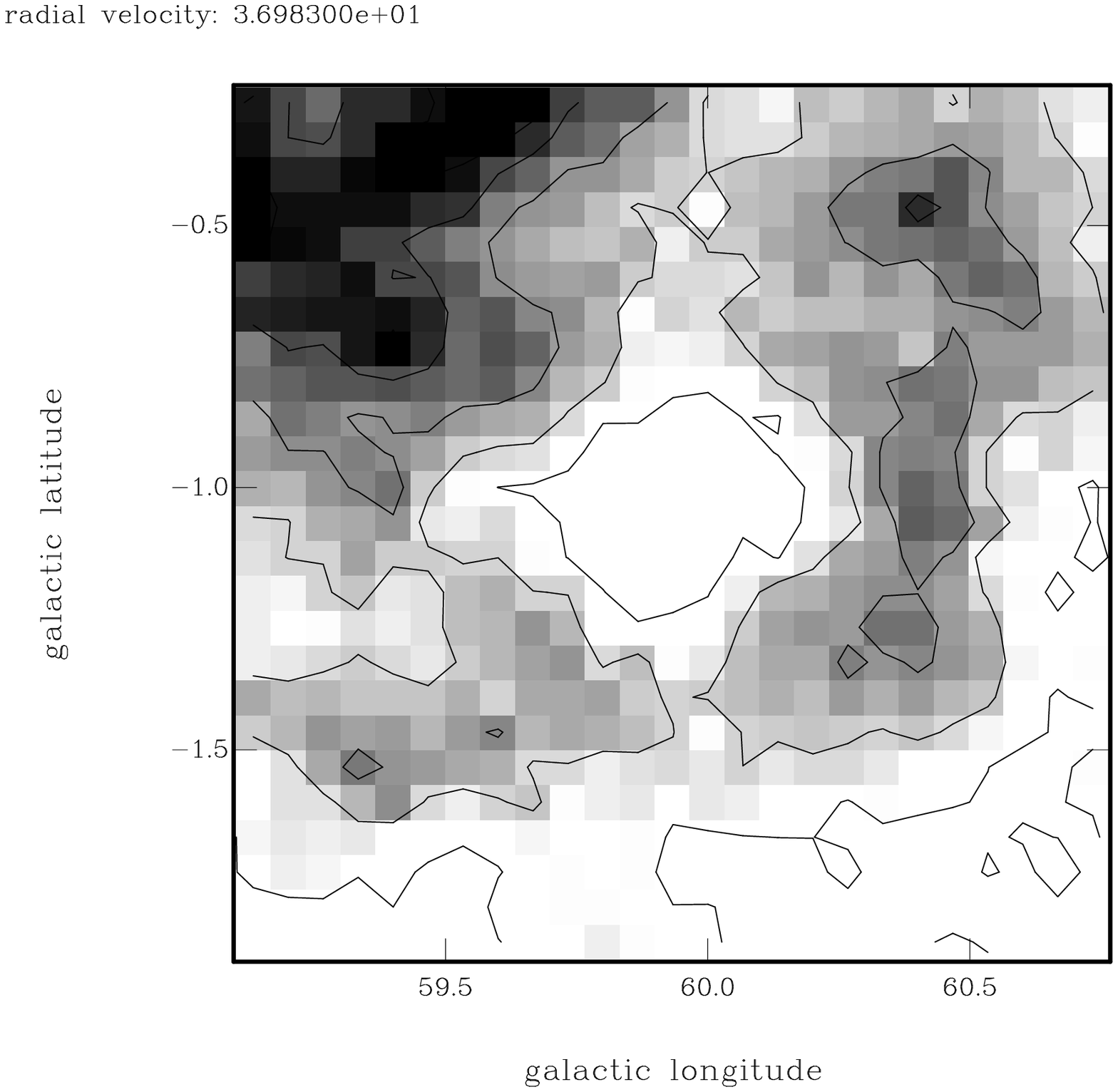}
 \includegraphics[width=8cm]{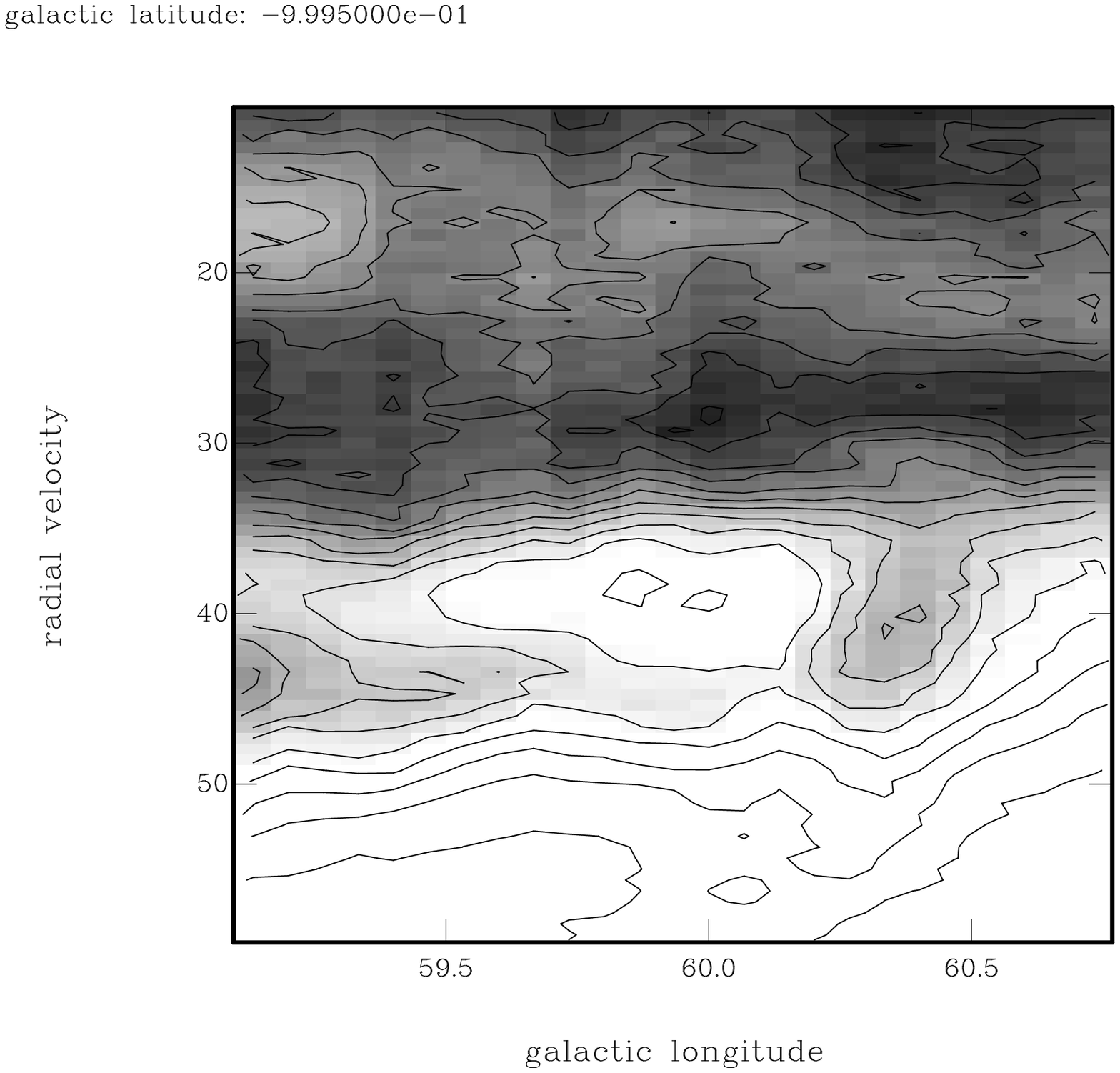}
 \caption{The shell GS59.9-1.0+38 in a velocity channel 
    ($v_{\mathrm{LSR}} = +37.0\ \mathrm{km s^{-1}}$) and the $lv$ cut
    ($b=-1.0^{\degr}$). The grey scale goes from white (the lowest
    temperature) to black (the highest temperature). The range of 
    temperatures in the left panel is (22.9 K, 61.5 K), contour
    values are 26.8 (1.9) 57.7 K; in the right panel the range is 
    (0.5 K, 88.8 K), contour values are 9.3 (4.4) 80.0 K.}
 \label{fig3}
\end{figure*}

GS59.9-1.0+38 is a small, probably young, spherical structure
(see Fig. \ref{fig3}). Both the approaching and receding hemisphere 
and a change of the shell diameter as a function of the radial 
velocity are seen. The shell is located in the inner Galaxy at the 
terminal velocity. Its dimensions are $60 \times 52$ arcmin, 
corresponding to a radius of roughly 35 pc; its expansion velocity 
is 9 $\mathrm{kms}^{-1}$. The right panel of Fig. \ref{fig3} may 
indicate a lower expansion velocity: a large gradient
of the brightness temperature with the velocity in this area hides
a part of the approaching hemisphere of the shell. Individual channel
maps or the $\Delta T_{\mathrm{B}}$ graph --- see Fig. \ref{fig2}
or Ehlerov\'a (2000) --- show the higher expansion velocity.

Quantities derived from the $\Delta T_B$ graph (see Fig. \ref{fig2}) 
are summarized in Table \ref{table1}.
FWHM gives the width of the line (if the line profile is Gaussian,
the dispersion $\sigma \simeq 0.6\ \mathrm{FWHM}$); $\mathrm{N_{\hi}}$ is
the column density of \hi swept up into the wall (or missing in the hole);
$m_{\mathrm{\hi}}$ is the derived mass of \hi swept up in walls (or missing
in the hole), assuming the radius of the shell to be 35 pc.

The velocity dispersion of the gas swept into the shell is quite small 
(1.5--2.0 $\mathrm{km s^{-1}}$) which is in agreement with the expected
high cooling rate in dense walls. 

The masses derived from walls and a hole are not the same,
but this is not very surprising, given the method and uncertainties
in deriving the background (see the section ``$\Delta T_B$ graphs''). 
As a reasonable estimate we adopt the value of the total mass 
$m_{\mathrm{tot}}$
($m_{\mathrm{tot}}=m_{\hi}/0.7$, where 0.7 is the solar abundance of
hydrogen):
$$m_{\mathrm{tot}} = 5.2 \times 10^3 M_{\sun}.$$ 
The corresponding volume density of the ISM at the position of
the ISM before the creation of the shell is
$$n_0 = 0.9\ \mathrm{cm}^{-3}.$$

The energy needed to create the structure is (from Eq. \ref{eq:cheval}):
$$E_{\mathrm{tot}} = 0.7 \times 10^{50}\mathrm{erg}.$$ 
This value is smaller than the "canonical"
energy of one supernova, $E_{\mathrm{SN}}=10^{51}\mathrm{erg}$.
This, in fact, is not unusual for observed shells, see e.g. Heiles
(1979).

The shell GS59.9-1.0+38 is probably young, from the analytical
solution (Sedov, 1959) we estimate its expansion age as $\simeq 1.5$ Myr.

\subsection{GS59.7-0.4+44}

\begin{table}[htb]
\begin{tabular}{lrrr}
\hline
~~ & FWHM & $N_{\mathrm{\hi}}$ & $m_{\mathrm{\hi}}$ \\
~~ & ($\mathrm{km s^{-1}}$) & ($\mathrm{cm^{-2}}$) & ($M_{\sun}$) \\
\hline
hole  & 8.4 & $2.7 \times 10^{20}$ & $4.0 \times 10^3$ \\
wall  & 7.7 & $1.2 \times 10^{20}$ & $1.0 \times 10^4$ \\
\hline
\end{tabular}
\caption{GS59.7-0.4+44:
FWHM is the width of the line, $\mathrm{N_{\hi}}$ is the column density 
of \hi swept up into the wall or missing in the hole derived from the
spectrum through the centre of the shell, $m_{\mathrm{\hi}}$ is the mass 
of \hi swept up in walls or missing in the hole, assuming the radius of 
30 pc. Only the receding wall is visible in velocity channels.}
\label{table2}
\end{table}

\begin{figure*}
 \includegraphics[width=8cm]{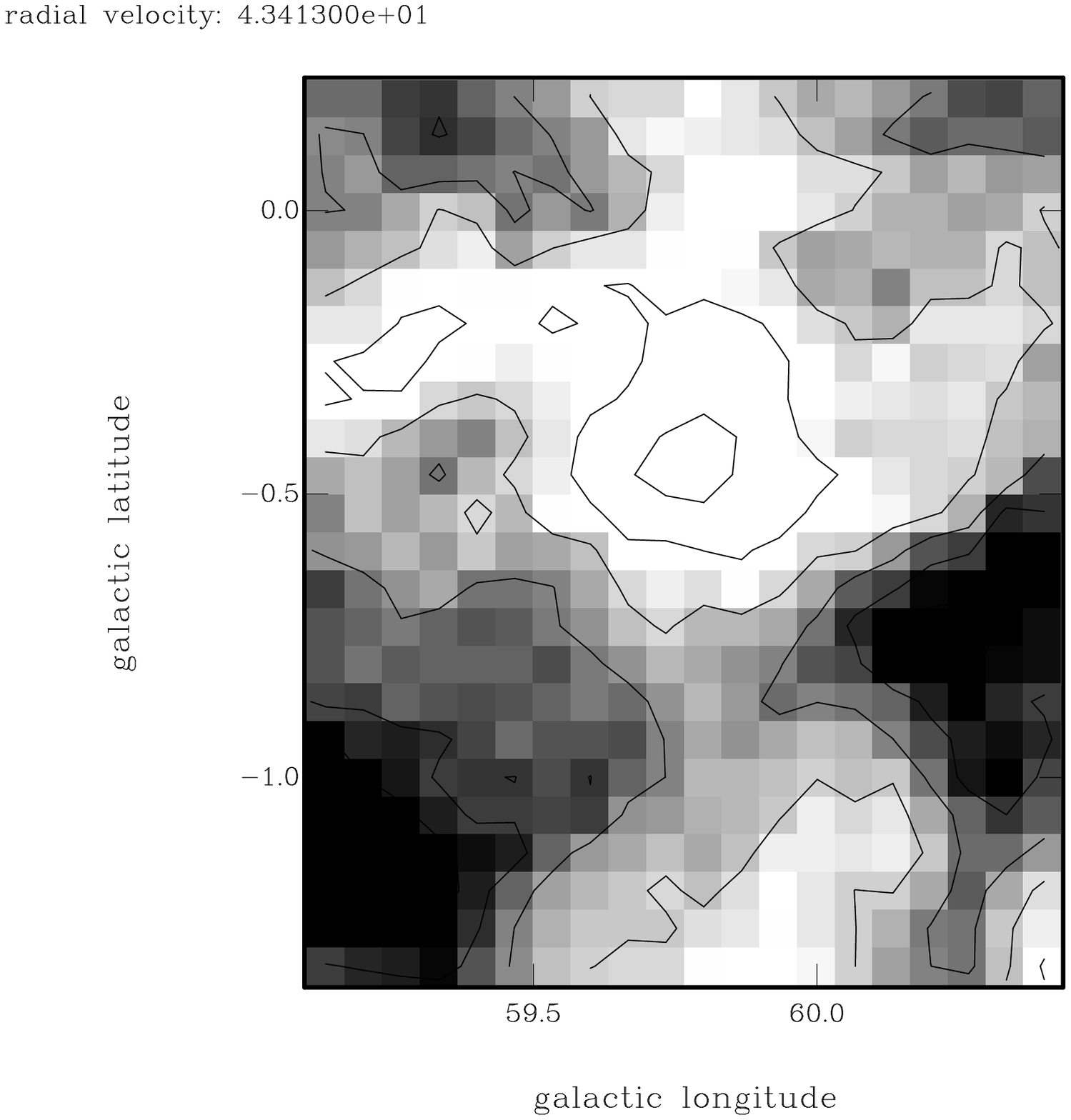}
 \includegraphics[width=8cm]{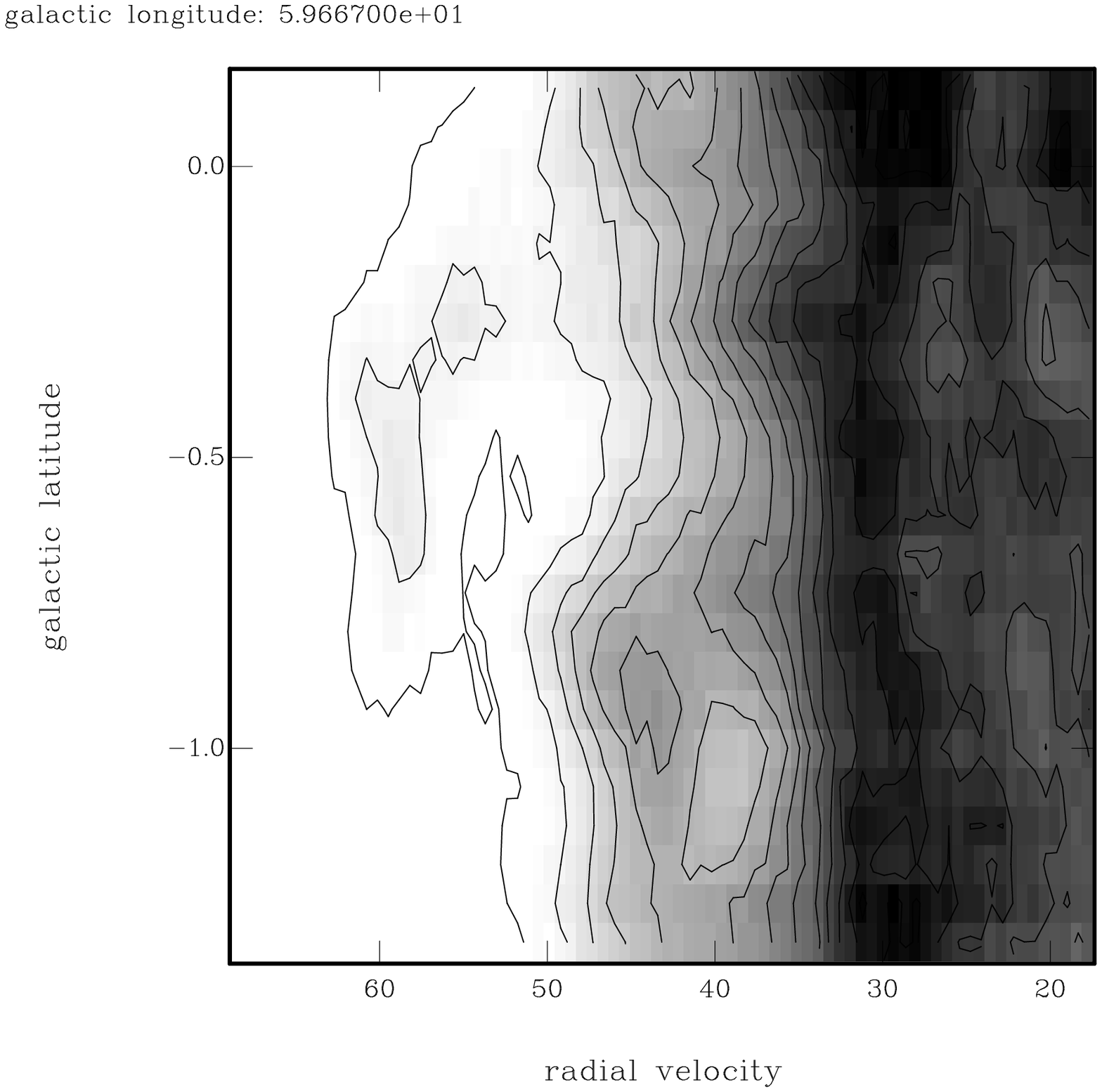}
 \caption{The shell GS59.7-0.4+44 in a velocity channel 
    ($v_{\mathrm{LSR}} = +43.4\ \mathrm{km s^{-1}}$) and the $bv$ cut
    ($l=59.67^{\degr}$). The grey scale goes from white (the lowest
    temperature) to black (the highest temperature). The range of 
    temperatures in the left panel is (4.5 K, 51.6 K), contour values
    are 9.2 (2.4) 46.8 K; in the right panel the range is (0.0 K, 91.3 K),
    contour values are 7.7 (4.6) 82.0 K.}
 \label{fig4}
\end{figure*}

GS59.7-0.4+44 (see Fig. \ref{fig4}) is another small spherical structure, 
in fact it is nearly a twin of GS59.9-1.0+38.
Like GS59.9-1.0+38, GS59.7-0.4+44 lies close to the
tangential point. Its radius is 24 arcmin (30 pc), its expansion
velocity is 14 $\mathrm{kms}^{-1}$ (for the explanation of the seemingly
lower expansion velocity in the $bv$ diagram see the previous section).
Table \ref{table2} summarizes observed properties of the shell.

A reasonable mass estimate is 
$$m_{\mathrm{tot}} = 7.5 \times 10^3 M_{\sun}$$ 
corresponding to
$$n_0 = 2.1\ \mathrm{cm}^{-3}.$$

Chevalier's estimate of the energy needed to create the  shell
GS59.7-0.4+44 is
$$E_{\mathrm{tot}} = 2.0 \times 10^{50}\mathrm{erg}.$$
As in the case of GS59.9-1.0+38 this energy is smaller than the
canonical value, but we suppose that the shell
could have been created by one supernova explosion.

The age of the shell is small, only about 1 Myr.

\subsection{GS061+00+51}

\begin{figure*}
 \includegraphics[width=8cm]{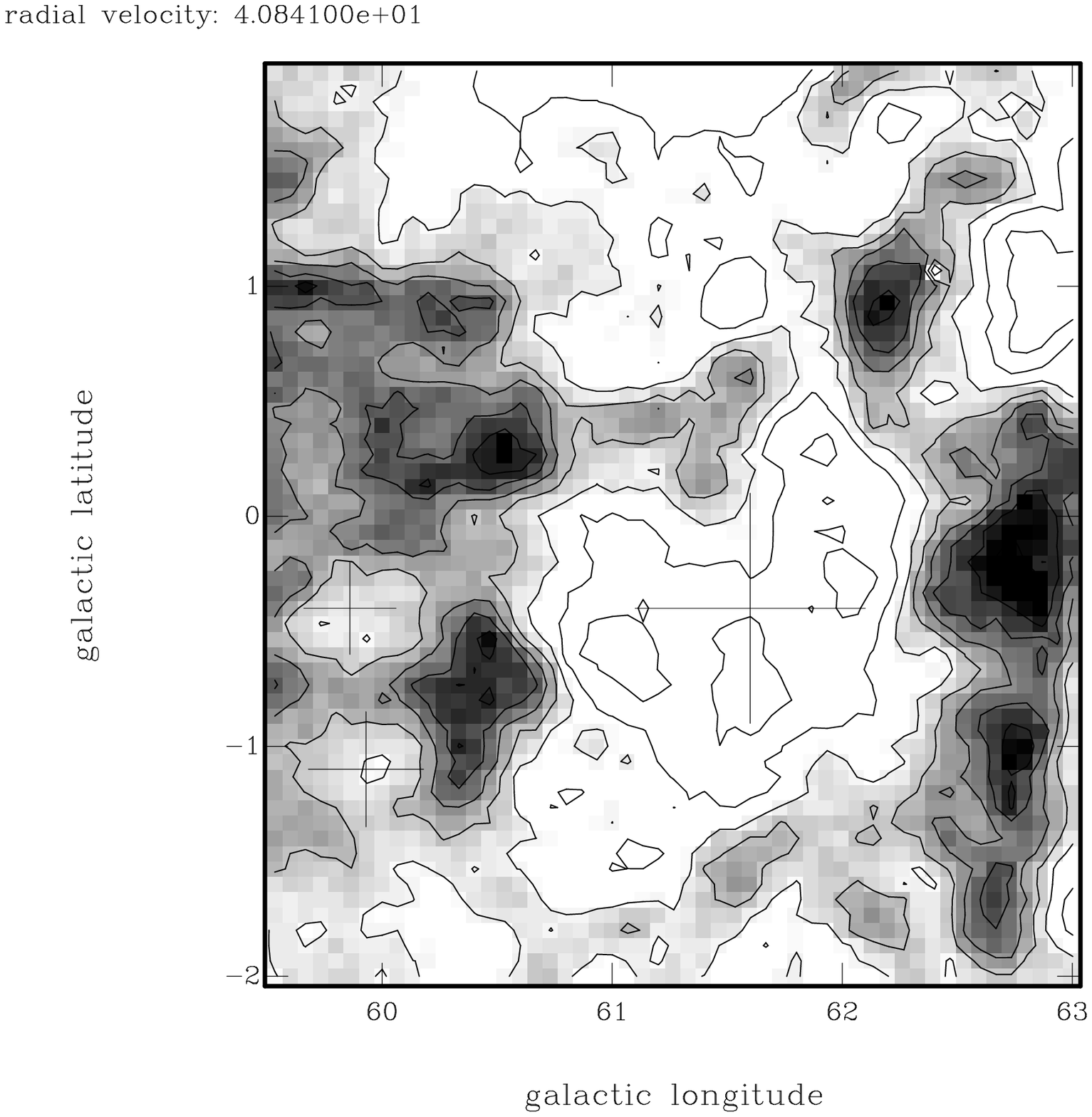}
 \includegraphics[width=8cm]{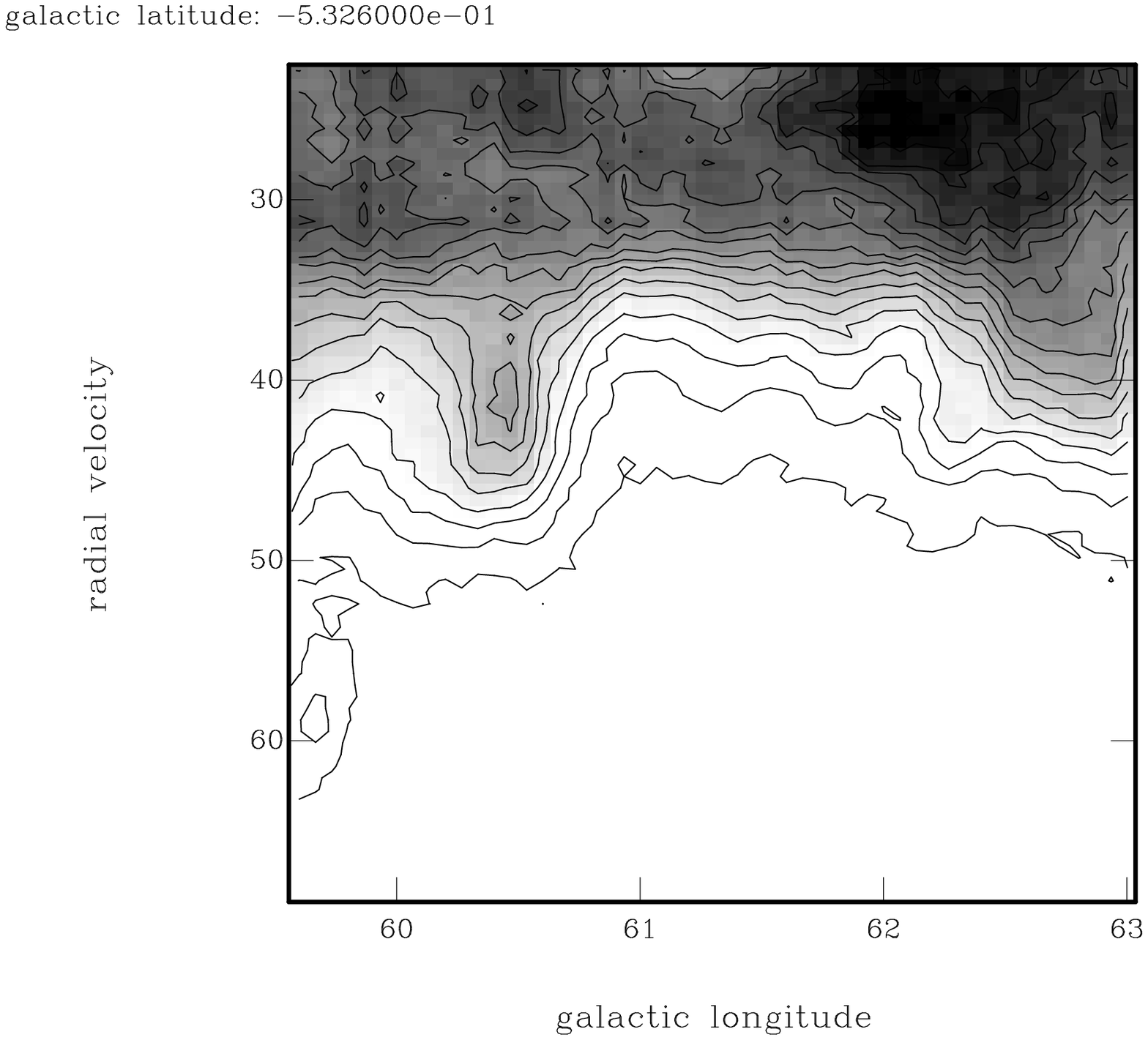}
 \caption{The shell GS061+00+51 in a velocity channel 
    ($v_{\mathrm{LSR}} = +40.8\ \mathrm{km s^{-1}}$) and the $lv$ cut
    ($b=-0.53^{\degr}$). Three crosses in the left panel show centers 
    of \hi shells GS061+00+51, GS59.9-1.0+38 and GS59.7-0.4+44.
    The grey scale goes from white (the lowest temperature) to black 
    (the highest temperature). The range of temperatures in the left
    panel is (7.0 K, 66.9 K), contour values are 13.0 (3.0) 61.0 K;
    in the right one panel the range is (0.0 K, 102.6 K),
    contour values are 8.7 (5.2) 92.2 K.}
 \label{fig5}
\end{figure*}

This is one of Heiles' expanding shells (Heiles, 1979), the only complete 
shell in our field which was known before. Its first detection and 
description can be found in Katgert, 1969. A receding part of the shell 
is not visible. The approaching hemisphere is seen clearly, and is quite
interesting. It is not a classical elliptical  shell, but, especially at 
lower velocities, it resembles a crescent (see Fig. \ref{fig5}). We can think 
of two possibilities to explain this shape:
\begin{enumerate}
\item{{\bf Merging theory:} \newline
   The shell was created by the merging of 3 (or more) small
   bubbles, each of them created by a stellar wind of one star or
   one supernova explosion. The bubbles interacted, but their untouched
   parts continued to expand, which led to a kind of "circularization". 
   Apart from the shape this theory also explains the denser walls inside 
   the hole (they are not seen very well in the printed version of the 
   picture).}
\item{{\bf Cloud theory:} \newline
   An originally quite regular shell encountered during the
   expansion a high density cloud. The cloud distorted
   the shape of the shell; at the places, where the shell emerged
   from the cloud, the expansion tends to make it spherical.
   Alternatively, the explosion (which created the shell) took place
   off center in a high density cloud, with a subsequent
   asymmetrical expansion.}
\end{enumerate}

GS061+00+51 is found near the terminal velocity. Its dimensions are
$\sim 2.4{\degr}$ in the $l$ direction and $2.9{\degr}$ in the $b$
direction, corresponding to 169 pc and 202 pc, respectively
(depending on how you define the edge of the shell).
Its expansion velocity is roughly 13 $\mathrm{kms}^{-1}$, only one hemisphere
is seen, in $\Delta T_{\mathrm{B}}$ graphs no reliable walls are identified.

The properties of the shell are:
\vskip0.2cm

\begin{tabular}{l}
$N_{\mathrm{\hi}} = 3.7 \times 10^{20} \mathrm{cm^{-2}}$ \\
$m_{\mathrm{tot}} = 1.0 \times 10^{5} M_{\sun}$ \\
$n_0 = 1.0\ \mathrm{cm^{-3}}$ \\
$E_{\mathrm{tot}} = 2.6 \times 10^{51} \mathrm{erg}$. \\
\end{tabular}

The energy, which created the shell, was probably released in
one or more supernova explosions.
The estimate of the shell age is $\tau \simeq 4$ Myr.

The dimensions of the structure as given by Heiles (1979) are slightly
higher than our values, which is caused by 1) the fact, that the
resolution of the Effelsberg radiotelescope is higher than that of the survey
in which Heiles identified the shells: viz. the \hi survey of 
Weaver \& Williams (1973) with a spatial resolution of 
$36\ \mathrm{arcmin}$ and a velocity resolution of 2 $\mathrm{kms}^{-1}$; 
and 2) uncertainties in defining the precise boundaries of the shell --- 
while there is no doubt about the existence and position of the structure, 
it is not completely clear, if all adjoining depressions belong to it.

Obviously (see Fig. \ref{fig5}), shells GS061+00+51, GS59.9-1.0+38
and GS59.7-0.4+44 are neighbours. 
GS061+00+51 is older and bigger than the other two, but not old
enough to trigger secondary star formation in the walls, which could
result in the creation of new small shells on the rim of the
old structure. We may be witnessing propagating star formation in one cloud
(or a cloud complex) which started at higher galactic longitudes 
and propagates toward the lower longitudes. The difference in ages
of GS061+00+51, GS59.9-1.0+38 and GS59.7-0.4+44 is about 3-4 Myr,
which suggests that the speed of the shock front compressing the gas
and triggering the star formation is around 40 $\mathrm{kms}^{-1}$
(this is a lower limit since we do not take into account the 
differences in radial distances). In a few million years the three bubbles 
should merge.

\subsection{GS62.1+0.2-18}

\begin{table}[htb]
\begin{tabular}{lrrr}
\hline
~~ & FWHM & $N_{\mathrm{\hi}}$ & $m_{\mathrm{\hi}}$ \\
~~ & ($\mathrm{km s^{-1}}$) & ($\mathrm{cm^{-2}}$) & ($M_{\sun}$) \\
\hline
hole  & 10.3 & $3.7 \times 10^{20}$ & $7.4 \times 10^4$ \\
wall  & 5.1 & $1.9 \times 10^{20}$ & $2.3 \times 10^5$ \\
\hline
$m_{\mathrm{tot}}$  & $1.6 \times 10^5$    & $M_{\sun}$         & \\
$n_0$               & 0.9                  & $\mathrm{cm^{-3}}$ & \\
$E_{\mathrm{tot}}$ & $3.6 \times 10^{51}$ & $\mathrm{erg}$     & \\
\hline
\end{tabular}
\caption{GS62.1+0.2-18:
FWHM is the width of the line, $\mathrm{N_{\hi}}$ is the column density 
of \hi swept up into the wall or missing in the hole, $m_{\mathrm{\hi}}$ 
is the mass of \hi swept up in walls or missing in the hole, assuming the 
radius of 110 pc. Only one wall is seen reliably in velocity channels. 
$m_{\mathrm{tot}}$ and $E_{\mathrm{tot}}$ 
are the best estimates of the total mass of the structure and the energy
needed for its creation, $n_0$ is the corresponding ISM density at the
position of the shell.}
\label{table3}
\end{table}

\begin{figure*}
 \includegraphics[width=8cm]{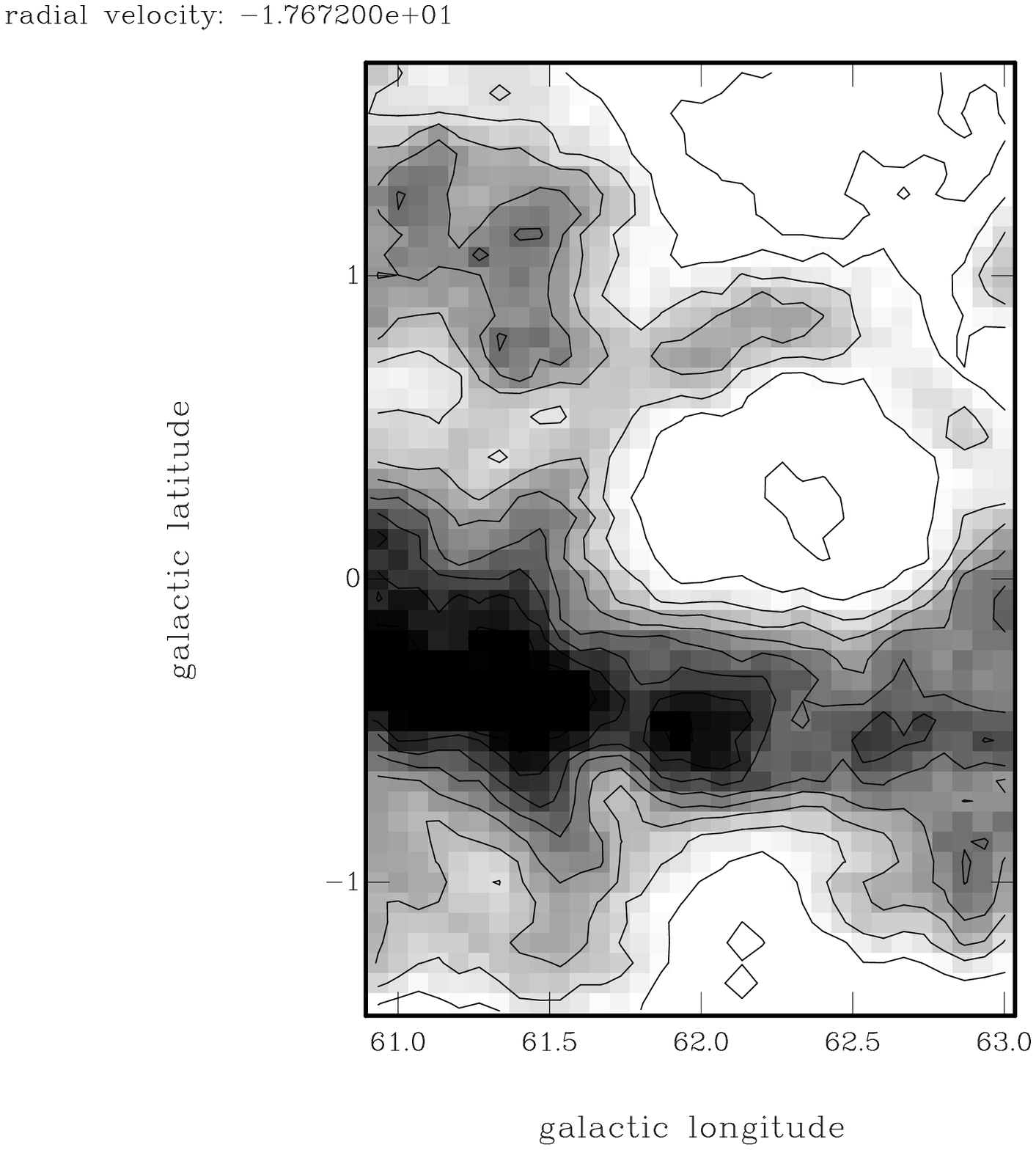}
 \includegraphics[width=8cm]{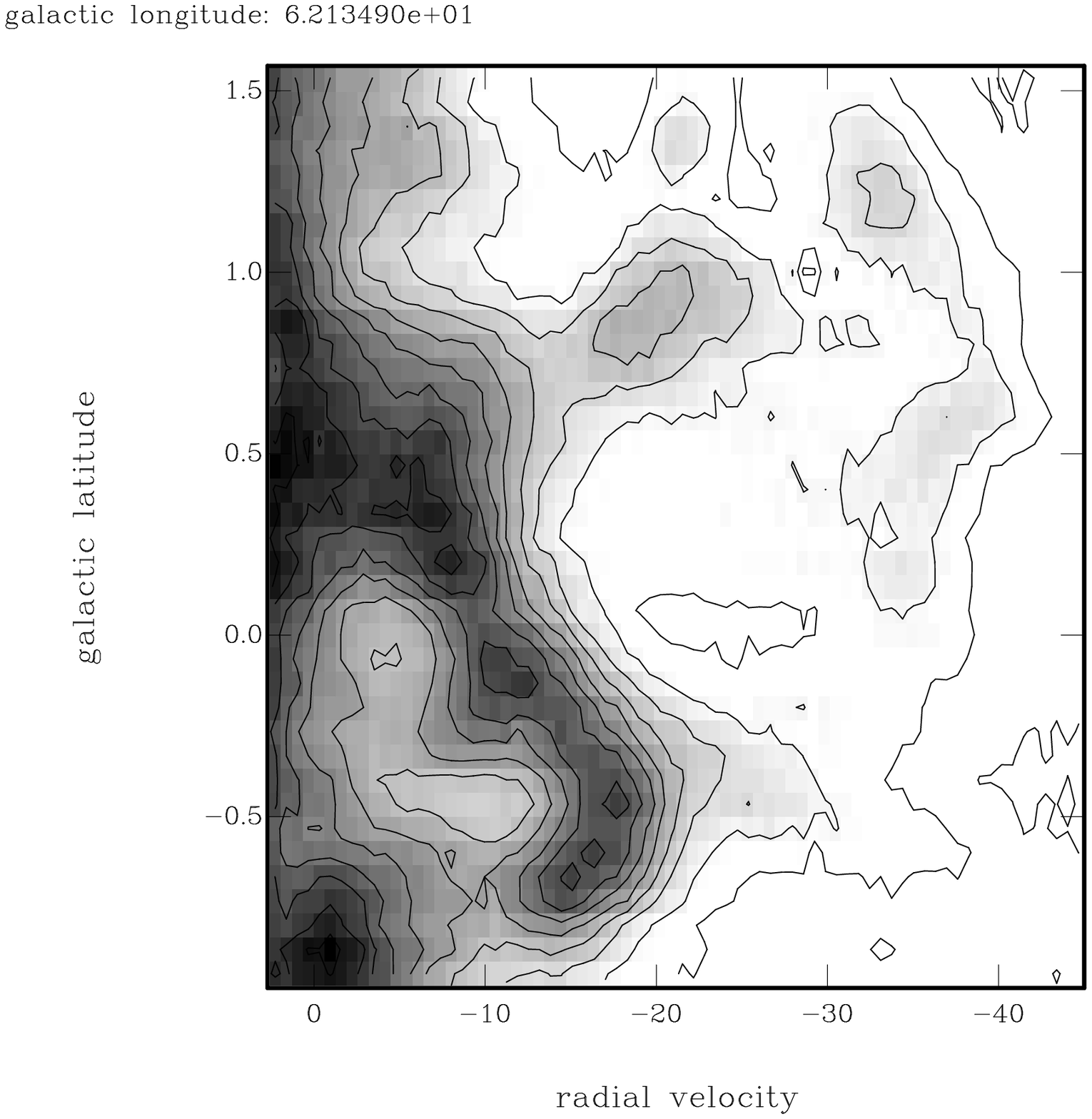}
 \caption{The shell GS62.1+0.2-18 in a velocity channel 
    ($v_{\mathrm{LSR}} = -17.7\ \mathrm{km s^{-1}}$) and the $bv$ cut
    ($l=62.13^{\degr}$). The grey scale goes from white (the lowest
    temperature) to black (the highest temperature). The range of 
    temperatures in the left panel is (7.3 K, 79.1 K), contour values
    are 14.5 (3.6) 72.0 K; in the right panel the range is (3.7 K, 92.9 K),
    contour values are 12.6 (4.5) 84.0 K.}
 \label{fig6}
\end{figure*}

This is a comparatively spherical shell in the outer Galaxy (see Fig. 
\ref{fig6}); it lies at a distance of 9.6 kpc.
Its radius is 40 arcmin, or $\sim 110\ \mathrm{pc}$ 
(in the $l$ direction it is 120 pc, 100 pc in the $b$ direction). 
The expansion velocity is 13 $\mathrm{kms}^{-1}$. Only one wall is
seen reliably. Table \ref{table3} summarizes properties of the shell.

One or more probably several supernovae were needed to create the shell 
GS62.1+0.2-18; its age is $\sim 5$ Myr.

\subsection{GS60.0-1.1-54}

\begin{table}
\begin{tabular}{lllrr}
\hline
$l$ & $b$ & ~~ & $N_{\mathrm{\hi}}$ & FWHM \\
~~  & ~~  & ~~ & $\mathrm{cm^{-2}}$ & $\mathrm{km s^{-1}}$ \\
\hline
$60.0^{\degr}$ & $+0.2^{\degr}$ & hole & $2.6 \times 10^{20}$ & 8.3 \\
~~            & ~~            & wall & $4.5 \times 10^{19}$ & 5.0 \\
$60.0^{\degr}$ & $-0.5^{\degr}$ & hole & $3.6 \times 10^{20}$ & 11.0 \\
~~            & ~~            & wall 1 & $2.4 \times 10^{19}$ & 2.7 \\
~~            & ~~            & wall 2 & $9.3 \times 10^{19}$ & 5.5 \\
$60.1^{\degr}$ & $-1.1^{\degr}$ & hole & $5.1 \times 10^{19}$ & 7.8 \\
~~            & ~~            & wall & $3.9 \times 10^{20}$ & 8.2 \\
\hline
\end{tabular}
\caption{GS60.0-1.1-54: 
$l$ and $b$ are coordinates of the spectrum taken through the structure,
FWHM is the width of the line, $\mathrm{N_{\hi}}$ is the column density 
of \hi swept up into the wall or missing in the hole.}
\label{table4}
\end{table}

\begin{figure*}
 \includegraphics[width=8cm]{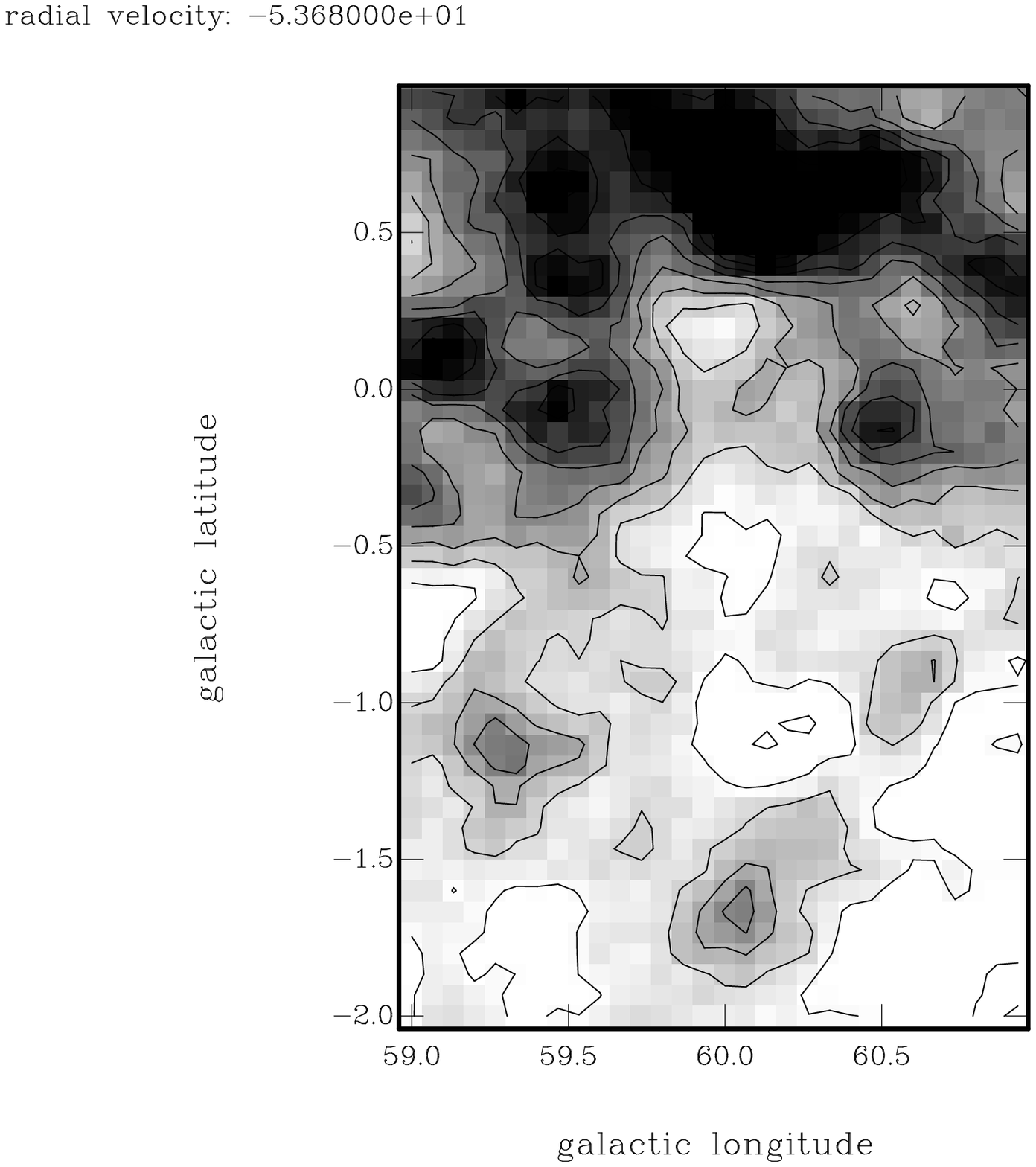}
 \includegraphics[width=8cm]{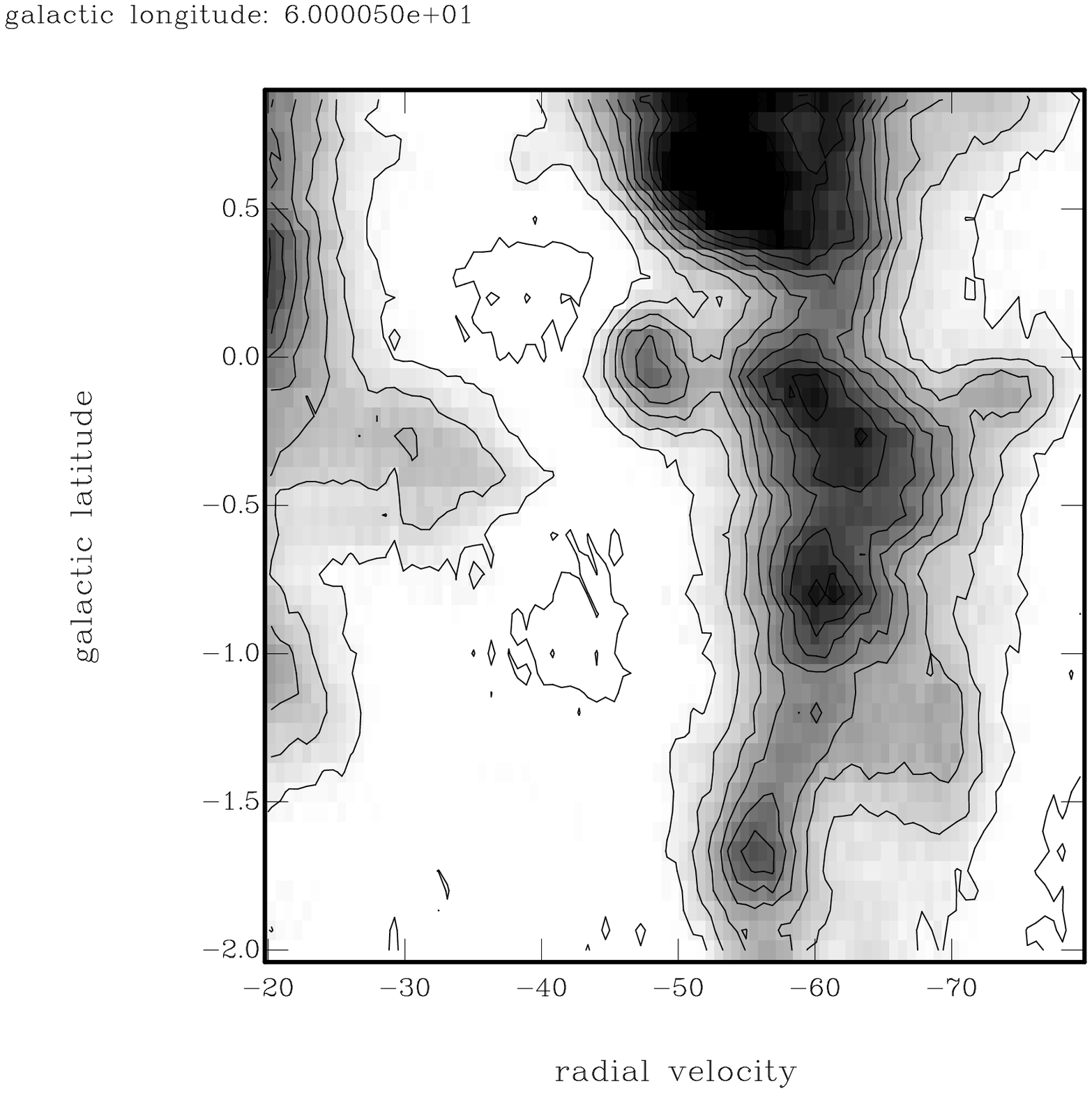}
 \caption{The shell GS60.1-1.1-54 in a velocity channel 
    ($v_{\mathrm{LSR}} = -53.7\ \mathrm{km s^{-1}}$) and the $bv$ cut
    ($l=60.0^{\degr}$). The grey scale goes from white (the lowest
    temperature) to black (the highest temperature). The range of 
    temperatures in the left panel is (6.3 K, 59.4 K), contour values
    are 11.6 (2.7) 54.0 K; in the right panel the range is (2.4 K, 56.9 K), 
    contour values are 7.8 (2.7) 51.5 K.}
 \label{fig7}
\end{figure*}

The shell GS60.0-1.1-54 is a highly non-spherical structure (Fig. \ref{fig7}).
It consists of a roughly spherical hole centered on ($60{\degr},0.2{\degr}$),
connected with a cone which opens to the halo, closed by an arc.
The shell lies in the outer Galaxy, at a distance of 13.7 kpc.
Its dimensions are about $2{\degr}$ (500 pc) in the $b$-direction,
the maximum diameter in the $l$-direction is  $1.7{\degr}$ (400 pc).
Though it is quite extended in the $b$-direction, it is not
an object in the Koo et al. (1992) catalog of galactic worm
candidates.

The \hi shell GS60.0-1.1-54 is an irregular structure, however,
it is probably not unique in the Milky Way. Its shape and dimensions are
similar to the Aquila supershell (Maciejewski et al., 1996). 
For a possible scenario how to create such a structure compare
the rightmost panel of Fig. 3 in Korpi et al. (1999) showing 
results of MHD simulations. The structure shown resembles the observations 
quite well, both in shape and dimensions.

The shell GS60.0-1.1-54 does not show the approaching hemisphere,
i.e. it is open at one side (or the wall is negligible). The 
receding hemisphere is visible: the small ``hole'' around 
$(60.0^{\degr}, +0.2^{\degr})$ changes diameter as expected
from the expanding structure with an expansion velocity of 
9 $\mathrm{km s^{-1}}$. The spectrum through $(60.0^{\degr}, -0.5^{\degr})$
also shows the expansion (17 $\mathrm{km s^{-1}}$).
The different expansion velocities are quite consistent with the idea 
that the fastest deceleration of the shell takes place in the densest part of 
the Galactic disk. The blown-out part at high
latitudes changes shape and dimensions with velocity, though
not in a very regular way. The best estimate of the expansion velocity
is 9 $\mathrm{km s^{-1}}$. 
Table \ref{table4} gives the column densities in different positions inside
the shell.

The shell GS60.0-1.1-54 is very irregular and therefore
we have not estimated its energy, as this is very unreliable.

\section{\hi observations of \hii regions}

\begin{figure*}
 \includegraphics[width=8cm]{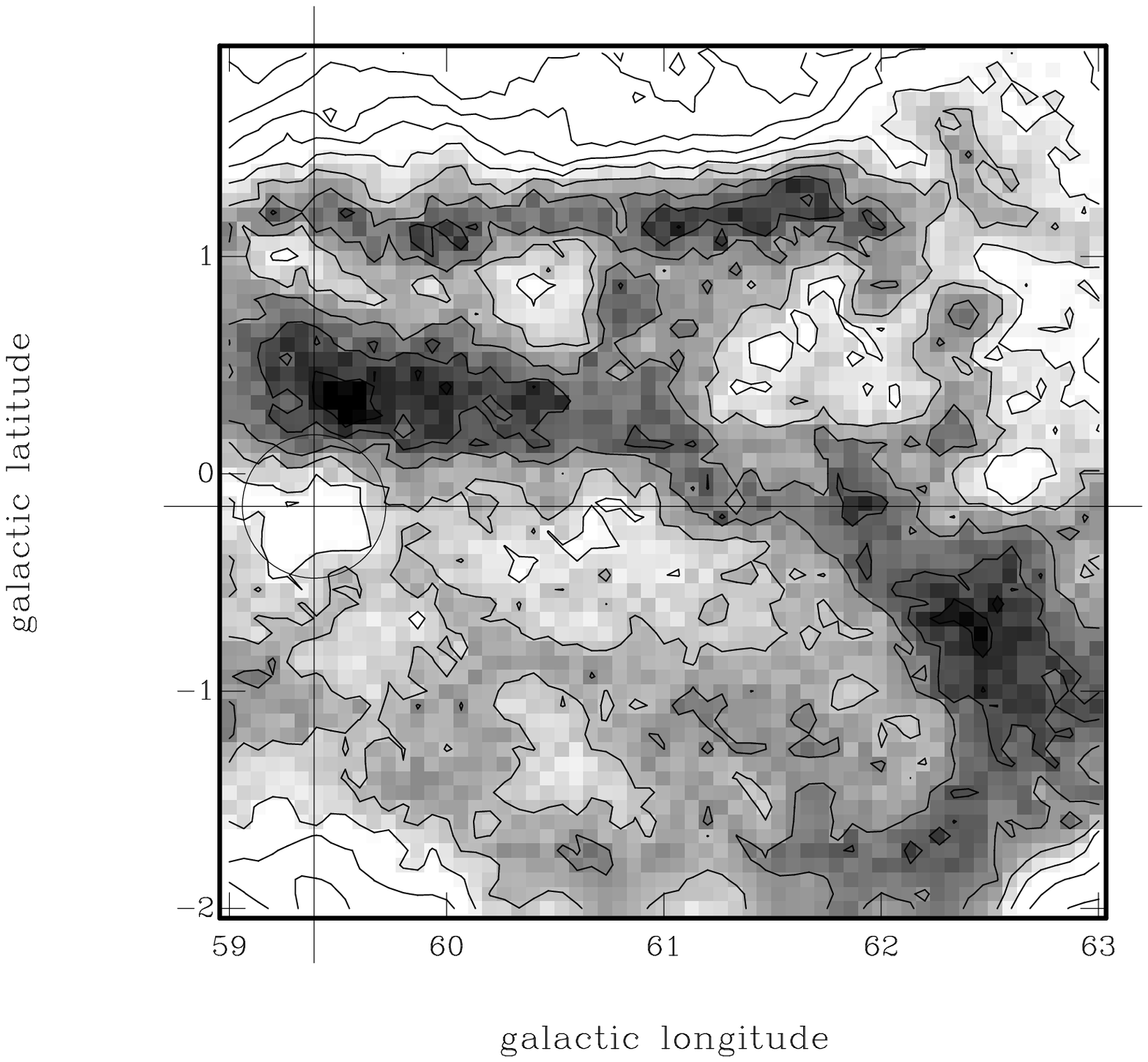}
 \includegraphics[width=8cm]{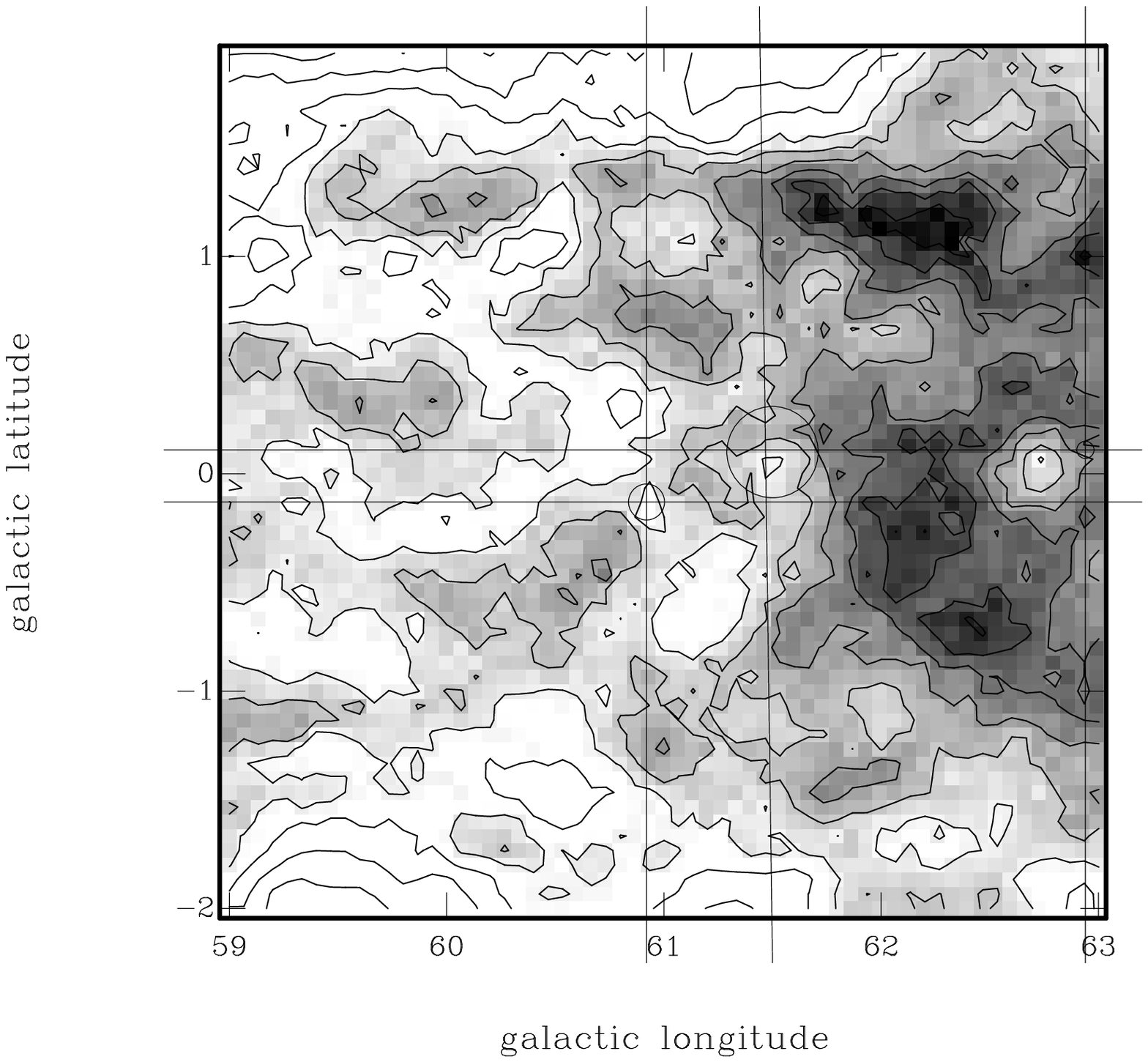}
 \caption{\hii regions in the observed field. The pixel map is the \hi
  column density between 
  $v_{\mathrm{LSR}} \in (+25,+33)\ \mathrm{kms}^{-1}$ with the
  S86 region (left panel); 
  \hi in $v_{\mathrm{LSR}} \in (+20,+25)\ \mathrm{kms}^{-1}$ 
  with regions (from left to right) S87, S88 and S89 (right panel).
  Positions of \hii regions are indicated by big crosses and circles
  corresponding to dimensions of regions (S89 is very small, with the
  diameter only barely exceeding the size of the pixel, it lies near to
  the edge of the map at the upper cross-section of the lines).
  The grey scale goes from white (the lowest column density)
  to black (the highest density). The range of 
  densities is in the left panel ($5.19 \times 10^{20} \mathrm{cm}^{-2}$, 
  $1.67 \times 10^{21} \mathrm{cm}^{-2}$), contour values are
  $6.34 \times 10^{20}$ ($1.15 \times 10^{20}$)
  $1.55 \times 10^{21} \mathrm{cm}^{-2}$;
  in the right panel the range is ($3.75 \times 10^{20} \mathrm{cm}^{-2}$, 
  $1.18 \times 10^{21} \mathrm{cm}^{-2}$), contour values are
  $4.56 \times 10^{20}$ ($8.05 \times 10^{19}$)
  $1.10 \times 10^{21} \mathrm{cm}^{-2}$.}
 \label{fig8}
\end{figure*}

There are four optical \hii regions in the observed field; S86,
S87, S88 and S89 (Sharpless, 1959). We examine the distribution
of the neutral hydrogen in their vicinity. Table \ref{table5} gives 
the properties of the four regions taken from Blitz \& Fich (1982).

\begin{table}[htb]
\begin{tabular}{lllll}
\hline 
name &   $l$ &   $b$ & $\mathrm{v_{CO}}$ ($\mathrm{km s^{-1}}$) & 
 $d$ (arcmin) \\
\hline
S86  & $59.39^{\degr}$ & $-0.15^{\degr}$ & 26.8 & 40  \\
S87  & $60.92^{\degr}$ & $-0.13^{\degr}$ & 22.7 & 10  \\ 
S88  & $61.49^{\degr}$ &  $0.10^{\degr}$ & 22.9 & 25  \\
S89  & $62.94^{\degr}$ &  $0.11^{\degr}$ & 25.6 &  5  \\
\hline
\end{tabular}
\caption{\hii regions in the observed field.
$l$ and $b$ are the coordinates of the centre, $v_{\mathrm{CO}}$ is the 
line-of-sight velocity of the CO counterpart of the optical region, 
$d$ is the diameter of the region.}
\label{table5}
\end{table}

\subsection{S86}

This \hii region is associated with the Vul OB1 association
(NGC 6823). At the position of S86 there is a clear hole in 
the \hi distribution, visible between 
$v_{\mathrm{LSR}} \in (+25,+31)\ \mathrm{kms}^{-1}$.
The \hii region lies inside the hole, its dimensions are comparable
to dimensions of the hole (see Fig. \ref{fig8}). The hole is
stationary.

The coincidence of the \hi hole and the \hii region is consistent with
the idea, that most gas in the vicinity of the OB association is
ionized and therefore not observed in 21 $\mathrm{cm}$ emission. 
The hole does
not expand, which may mean, that no SN has exploded so far in the
cluster (which is consistent with age estimates of NGC 6823: 2-7 Myr;
Massey et al., 1995).

\subsection{S87}

S87 is a source observed in optical, infrared, radio recombination
lines (RRL) and molecular line emission (Barsony, 1989; 
Onello et al. 1991). It has
a compact core surrounded by an extended structure oriented south-east
(i.e. perpendicular to the galactic plane). It interacts
with a molecular cloud.

The \hii region S87 lies inside the hole in the \hi distribution, visible
between $v_{\mathrm{LSR}} \in (+20,+26)\ \mathrm{kms}^{-1}$. Again, this hole 
is stationary.

\subsection{S88}

S88 is also observed in RRLs, molecular line emission, infrared and
optical (Wood \& Churchwell, 1989; Onello et al., 1991). The region has an
ultracompact core with a complex, multi-peaked structure.

S88 probably lies at the boundary between a dense sheet of gas and a more
rarefied medium. At the position and the radial velocity of the region
there is a small hole visible in a few velocity channels around 
$v_{\mathrm{LSR}} = 23\ \mathrm{kms}^{-1}$, but definitely not as 
pronounced as in the case of S86 or S87. This hole is a part of the bigger 
empty region (see Fig. \ref{fig8}).

\subsection{S89}

S89 lies in a dense region (see Fig. \ref{fig8}). It is not
situated inside any hole, at least not in the predicted velocity range,
but it lies just on the edge of a small hole, visible between
$v_{\mathrm{LSR}} \in (+20,+25)\ \mathrm{kms}^{-1}$. The physical association of these
two structures, an \hii region and an \hi hole, is unclear, but cannot
be excluded.

\subsection{Summary}

In two out of four cases (S86, S87) we find a clear trace of the Str\"omgren
sphere in the \hi distribution, i.e. a stationary hole. In one case (S88)
the connection \hii region -- \hi hole was not very obvious --- there are
depressions at the position of the region, but nothing really convincing.
Maybe simply the gas distribution in the vicinity of S88
is so chaotic, that the nice Str\"omgren sphere does not exist. The
region S89 does not lie inside a hole, but on the edge of one.

The chance coincidence of unrelated \hii regions and \hi holes cannot be
excluded, because of the distance ambiguity, but at least for S86 and S87
the probability of this coincidence is small, as not only the positions and
radial velocities, but also the dimensions of \hii regions and \hi holes agree.

The area where all these \hii regions lie, i.e. 
$v_{\mathrm{LSR}} \in (+10,+30)\ \mathrm{kms}^{-1}$, 
is a very turbulent region, full of structures on many
scales (in Ehlerov\'a, 2000, it was described as a strange kind of
a complex, multicomponent \hi shell
GS60.1-0.3+15). This is partly the reason why none of the 
\hi holes mentioned was identified as an independent \hi shell.

\section{Discussion --- Summary}

The $4{\degr} \times 4{\degr}$ field contains a rich variety
of structures. Due to its limited size, selection effects play heavily
against any statistical or general considerations and we can only
describe individual structures.

\begin{itemize}
 \item{We reidentified Heiles' supershell GS061+00+51. We confirm
       the previously 'detected' properties (size, expansion velocity),
       though its morphology is more complex than was known before.
     }
 \item{We found two young spherical expanding shells GS59.9-1.0+38
       and GS59.7-0.4+44 (with $R_{\mathrm{sh}} \sim 30\ \mathrm{pc}$
       and $v_{\mathrm{exp}} \sim 10\ \mathrm{km s^{-1}}$). In the
       future they might merge with one another (in about 1 Myr)
       and later (3-4 Myr) with a bigger shell GS061+00+51 which is 
       close by.
      }
 \item{Another regular elliptical shell GS62.1+0.2-18 was identified 
       in the outer Galaxy.
      }
 \item{An irregular shell GS60.0-1.1-54 was found. This shell blows out
       from the Galactic disk. This behaviour is expected from \hi
       supershells, but this structure lies in the outer Galaxy, where
       we do not expect any significant star formation and where the
       gaseous disk is thick.
      }
 \item{The initial density of the neutral hydrogen at positions of \hi 
      shells is in all cases higher than the average value; the typical
      $n_0$ is around 1 $\mathrm{cm^{-3}}$.
      }
\end{itemize}

Apart from the described \hi shells, which we consider to be non-turbulent
structures (i.e. created by something else than purely by the turbulence, most 
probably by the activity of massive stars), there are other
objects, which are less ``homogeneous'', such as 
holes in the \hi distribution connected with \hii regions (examples:
\hii regions S86 and S87). However, we are not always able to connect
the features in the \hi with features in \hii (S88 and maybe also S89).
Chaotic holes, arcs and sheets seen in our observations seem to be
turbulent structures with no direct connection to physical objects.

Summing up, it seems that there are two types of
``shell-like'' structures found in the \hi distribution. The first,
formed by consistent structures, that are coherent in the
position-velocity space, is less abundant than the second type, which
contains non-coherent objects. We believe that these second type structures
are created mainly due to the turbulence in the ISM.
We identify the first group of objects with structures
known as \hi shells, as they fulfill the usual criteria put on shells.
This is good news concerning the question about the existence of \hi
shells. The bad news is the fact that there is no well-defined boundary
between the two types of structures.

\begin{acknowledgements}
Authors gratefully acknowledge financial support by the Grant
Agency of the Academy of Sciences of the Czech Republic under the
grant No. A3003705/1997 and support by the grant project of the
Academy of Sciences of the Czech Republic No. K1048102.
SE would like to thank MPIfR for the hospitality during her stay in Bonn.
\end{acknowledgements}

{}

\end{document}